\documentclass[draftcls, onecolumn]{IEEEtran}

\usepackage{amsmath}
\usepackage{amssymb}
\usepackage{multirow}
\usepackage{array}
\usepackage{threeparttable}
\usepackage{enumitem}
\usepackage{graphicx}
\usepackage{float}
\usepackage{color}
\usepackage{cite}

\newtheorem{theorem}{Theorem}
\newtheorem{corollary}{Corollary}
\newtheorem{remark}{Remark}

\allowdisplaybreaks

\begin{document}
%
\title{A Relay Can Increase Degrees of Freedom \\ in Bursty Interference Networks}
%
%
%

\author{Sunghyun~Kim,~\IEEEmembership{Student,~IEEE,}
        I-Hsiang~Wang,~\IEEEmembership{Member,~IEEE,}
        and~Changho~Suh,~\IEEEmembership{Member,~IEEE}
\thanks{S. Kim and C. Suh are with Department
of Electrical Engineering, Korea Advanced Institute of Science and Technology, Daejeon, South Korea (e-mail: koishkim@kaist.ac.kr; chsuh@kaist.ac.kr).}
\thanks{I.-H. Wang is with Department of Electrical Engineering, National Taiwan University, Taipei, Taiwan (e-mail: ihwang@ntu.edu.tw).}
}

\maketitle

\begin{abstract}
We investigate the benefits of relays in multi-user wireless networks with bursty user traffic, where intermittent data traffic restricts the users to bursty transmissions. To this end, we study a two-user \emph{bursty} MIMO Gaussian interference channel (IC) with a relay,
where two Bernoulli random states govern the bursty user traffic. We show that an in-band relay can provide a degrees of freedom (DoF) gain in this bursty channel. This beneficial role of in-band relays in the bursty channel is in direct contrast to their role in the non-bursty channel which is not as significant to provide a DoF gain. More importantly, we demonstrate that for certain antenna configurations, an in-band relay can help achieve \emph{interference-free} performances with increased DoF. We find the benefits particularly substantial with low data traffic, as the DoF gain can grow \emph{linearly} with the number of antennas at the relay. In this work, we first derive an outer bound from which we obtain a necessary condition for interference-free DoF performances. Then, we develop a novel scheme that exploits information of the bursty traffic states to achieve them.
\end{abstract}


%
\IEEEpeerreviewmaketitle

\section{Introduction}
\label{sec:introduction}

\IEEEPARstart{I}{n} establishing the fundamental performance limits of communication systems, it has been a conventional assumption that transmitters send signals at all times. A rationale behind the assumption is that a sufficient amount of data to transfer is always available, thus continuous transmissions take place.

However, the assumption is not necessarily true in practice. 
Intermittent data traffic restricts the amount of data to transfer, thus {bursty} transmissions take place. In fact, it is burstiness that needs much attention to characterize the fundamental performance limits in more practical contexts.

The bursty behavior of transmitters leads to an interesting phenomenon when multiple nearby users communicate at the same time. Interference, which has been a major barrier to performance improvement in wireless systems, also becomes bursty. As growing availability of wireless systems inevitably brings about bursty interference, effective ways of harnessing bursty interference are in high demand.

Our primary interest is to see if in-band relays can play a key role in mitigating bursty interference and provide gains in multi-user wireless networks. A past work~\cite{jafar:it09} found the role of in-band relays pessimistic in interference networks as they provide no degrees of freedom (DoF) gain, but our work on bursty interference tells a different story.

From an observation in a simple single-user network, one can anticipate promising benefits that relays can offer in bursty networks. To see this, consider a \emph{bursty} MIMO relay channel (RC), where the transmitter sends signals with probability $p$; the transmitter has a large number of antennas; the receiver and the relay have 1 and $L$ respectively. From the standard cut-set argument, we obtain an outer bound on the DoF: $\min\{p(1+L), 1\}$, which is also achievable as the cut-set bound is well known to be tight in single-source single-destination networks~\cite{gamalkim:nit}. Observe that the DoF is $p$ without a relay $(L=0)$, and it becomes strictly greater with a relay $(L \geq 1)$. From the observation, we see that an in-band relay in the single-user bursty network can provide a DoF gain. Also, we see that the gain can grow linearly with $L$ in low-traffic regimes $(p \ll 1)$.

To give an intuitive explanation, the DoF gain comes from the receive-forward operation of the relay: it receives $L$ extra symbols when the transmitter is active and forwards them to the receiver in a first-in-first-out (FIFO) manner when the transmitter is inactive. The relay exploits idle moments of the transmitter, and provides the receiver with useful symbols even when the transmitter is inactive.

The benefits that a relay brings into the single-user bursty network motivates us to further examine if the benefits carry over into multi-user interference networks with bursty user traffic. Particularly interested in possibilities that a relay can mitigate bursty interference between multiple users to a great extent and offer the benefits to each user as if no interference is present, we ask: can relays play a significant role in {bursty} interference networks to help achieve \emph{interference-free} DoF performances?

To answer this question, we consider a two-user \emph{bursty} MIMO Gaussian interference channel (IC) with a relay, where two independent Bernoulli random processes govern the bursty data traffic of the users. We derive an outer bound from which we obtain a necessary condition on the antenna configuration for interference-free DoF performances. And we develop a novel scheme that harnesses information of the bursty traffic states. Through this information, our scheme enables the relay and the transmitters to cooperate in a beneficial manner, and provides a significant DoF gain over the channel without a relay. More importantly, the scheme reveals that a relay can indeed help achieve \emph{interference-free} performances with increased DoF. We find the presence of a relay particularly beneficial with low data traffic, as the DoF gain can grow \emph{linearly} with the number of antennas at the relay. Our results show that the role of relays in the bursty channel is crucial in contrast to their role in the non-bursty channel, as in-band relays cannot increase DoF in the non-bursty channel~\cite{jafar:it09}.

Considerable work has been done toward understanding ICs with relays. Numerous techniques developed in the IC and the RC have been combined and applied to ICs with various relay types. Although we focus on in-band relays, out-of-band relays have also been of interest.
Sahin-Simeone-Erkip~\cite{sahin:it11} and Tian-Yener~\cite{yener:it12} considered an IC with an out-of-band relay that receives and transmits in a band orthogonal to the IC. Sridharan et al.~\cite{sridharan:isit08} considered an IC with an out-of-band reception and in-band transmission relay.

In-band relays have drawn great attention.
Sahin-Erkip~\cite{sahin:07} first considered an IC with an in-band relay, and proposed an achievable scheme that employs a decode-forward scheme~\cite{Cover:it79} and rate-splitting~\cite{carleial}. The relay decodes common and private messages, and forwards them by allocating its power across them.
Mari{\'c}-Dabora-Goldsmith~\cite{goldsmith:it12} suggested a different decode-forward based achievable scheme. The relay intentionally forwards interfering signals to enhance the reception of interference at the receivers, so that interference cancellation can be facilitated.
Other works~\cite{yu:ita10, yener:it11, noisy:it11, kang:isit13, tobias:it14} mostly proposed achievable schemes that extend a compress-forward scheme~\cite{Cover:it79}. In the schemes, the relay forwards compressed descriptions of its received signals, and the receivers decode messages based on their own received signals and the descriptions.

A distinction of our work compared to all past works is that we investigate realistic scenarios where intermittent data traffic limits the users to bursty transmissions. Wang-Diggavi~\cite{wang:spawc13} first considered such scenarios and studied a bursty Gaussian IC without a relay. Extending the results in~\cite{wang:spawc13} to the MIMO channel with a relay, we demonstrate that relays can offer substantial benefits in the bursty MIMO Gaussian IC.


\section{Model}
\label{sec:model}
\begin{figure}[!t]
\centering
\includegraphics[width=0.7\columnwidth]{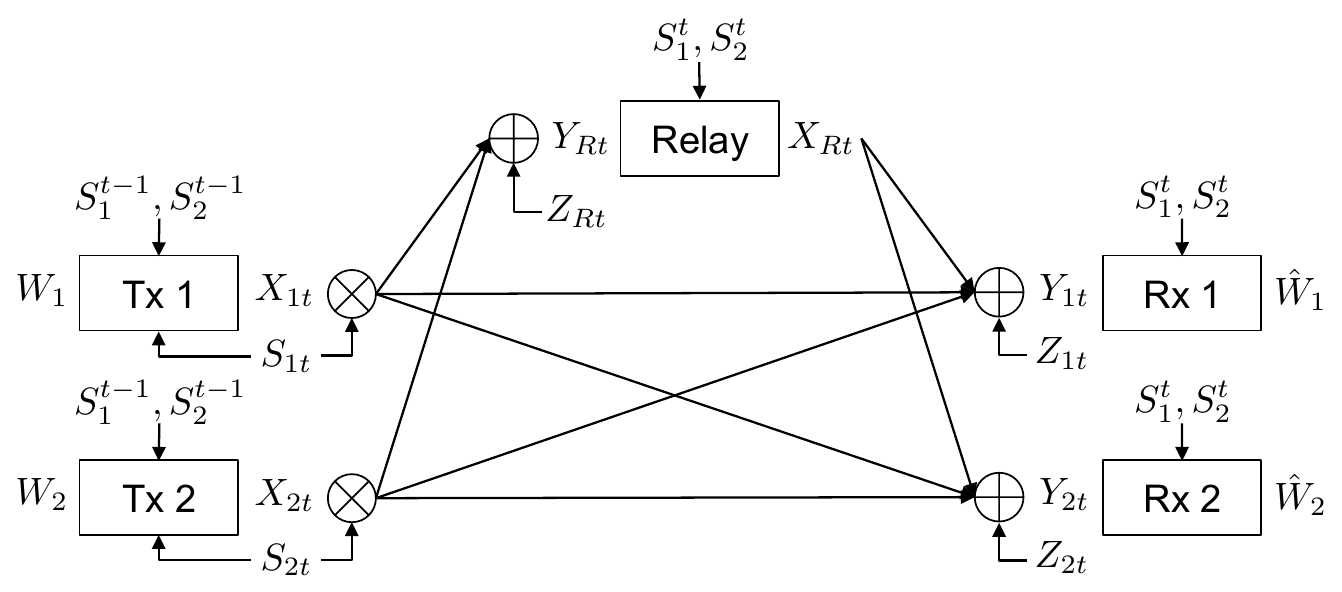}
\caption{Bursty MIMO Gaussian interference channel with an in-band relay.}
\label{fig:model}
\end{figure}

Fig.~\ref{fig:model} describes the bursty MIMO Gaussian interference channel (IC) with an in-band relay. The transmitters, the receivers, and the relay have $M$, $N$, and $L$ antennas, respectively. \mbox{Transmitter $k$} wishes to deliver \mbox{message $W_k$} reliably to \mbox{receiver $k$}, $\forall k = 1,2$. Let $X_{kt} \in \mathbb{C}^M$ be the encoded signal of \mbox{transmitter $k$} at \mbox{time $t$}, and $X_{Rt} \in \mathbb{C}^L$ be the encoded signal of the relay at \mbox{time $t$}. We introduce traffic \mbox{states $S_{kt}$} to govern bursty transmissions caused by intermittent data traffic that limits the amount of data to transfer at the transmitters\footnote{Finite-size buffers at nodes can be another source of burstiness, as they limit the amount of data available for transmission and reception. In this work, we consider intermittent data traffic to be a primary source of burstiness.}. We consider the case where the transmitters lack coordination in sharing a common communication medium\footnote{In wireless systems, distributed media access control protocols can lead to imperfect coordination of transmissions between multiple transmitting nodes.}. To capture uncoordinated bursty transmissions, we assume $S_{kt}$ to be independent, Bern($p$) and i.i.d. over time. The relay is not restricted to bursty transmissions. Unlike the transmitters, the relay has no intention to transfer its own data, which can be limited by intermittent data traffic that leads to bursty transmissions. {Rather, the relay generates its encoded signals based on its past received signals, so that it can help transmitter-receiver links communicate with better performance. Thus, the relay can send its signals at all times as long as it has past received signals.} Additive noise terms $Z_{kt}$ and $Z_{Rt}$ are assumed to be independent, $\mathcal{CN}(0,\mathbf{I}_N)$ and $\mathcal{CN}(0,\mathbf{I}_L)$, and i.i.d. over time. Let $Y_{kt} \in \mathbb{C}^N$ be the received signal of \mbox{receiver $k$} at \mbox{time $t$}, and $Y_{Rt} \in \mathbb{C}^L$ be the received signal of the relay at \mbox{time $t$}:
\begin{align*}
Y_{kt} & = \mathbf{H}_{k1} S_{1t} {X}_{1t} + \mathbf{H}_{k2} S_{2t} {X}_{2t} + \mathbf{H}_{kR} X_{Rt} + Z_{kt}, \\
Y_{Rt} & = \mathbf{H}_{R1} S_{1t} {X}_{1t} + \mathbf{H}_{R2} S_{2t} {X}_{2t} + Z_{Rt}.
\end{align*}
The matrices $\mathbf{H}_{ji}$, $\mathbf{H}_{Ri}$, and $\mathbf{H}_{jR}$ describe the time-invariant channels from \mbox{transmitter $i$} to \mbox{receiver $j$}, from \mbox{transmitter $i$} to the relay, and from the relay to \mbox{receiver $j$}, $\forall i,j = 1,2$. All channel matrices are assumed to be full rank.

We assume current traffic states are available at the receivers and the relay, since receiving nodes can detect which transmitting node is active, for instance, by measuring the energy levels of incoming signals. Also, we assume the transmitters get feedback of past traffic states from the receivers. {Through the feedback, the uncoordinated transmitters can devise ways to cooperate. Each transmitter knows its own current traffic state as it processes the arrivals of data for transmission}. Thus, \mbox{transmitter $k$} generates its encoded signal at \mbox{time $t$} based on its own message, its current traffic state, and the feedback of past traffic states:
\begin{gather*}
X_{kt} = f_{kt}(W_k, S_{kt}, S^{t-1}).
\end{gather*}
Shorthand notation $S_t$ stands for ($S_{1t}, S_{2t}$) and $S^{t-1}$ stands for the sequence up to $t-1$. The relay generates its encoded signal at \mbox{time $t$} based on its past received signals, and both past and current traffic states:
\begin{gather*}
X_{Rt} = f_{Rt}(Y_R^{t-1}, S^t).
\end{gather*}
Shorthand notation $Y_R^{t-1}$ stands for the sequence up to $t-1$. We can also consider the case where \mbox{transmitter $k$} knows its future traffic states non-causally, and generates its encoded signal at \mbox{time $t$} based on its own message, all of its own traffic states, and the feedback of past traffic states: $X_{kt} = f_{kt}(W_k, S_k^n, S_l^{t-1})$, where $l$ refers to the other transmitter. As a separate node, the relay has no knowledge of the future traffic states of the transmitters. Thus, the relay generates its encoded signals in the same way. We discuss the non-causal case in Section~\ref{sec:discussion}.

We define the DoF region as follows:
\begin{align*}
\mathcal{D} = \left\{ (d_1, d_2) : \begin{array}{l} \exists (R_1, R_2) \in \mathcal{C}(P) \text{ such that } d_k = \lim_{P \rightarrow \infty} \frac{R_k}{\log P} \end{array} \right\}.
\end{align*}
$\mathcal{C}(P)$ is the capacity region with power \mbox{constraint $P$} on each antenna. We follow the conventional way of defining the DoF regions of non-bursty channels. Another way to define the DoF regions of bursty channels would be to divide rate tuples in the capacity region by $p\log(P)$, to observe how the rate tuples scale in comparison to the capacity of a bursty Gaussian point-to-point channel at high signal-to-noise ratio ($P \rightarrow \infty$). However, when our main interest is to compare the DoF regions of the Gaussian IC with and without a relay, the new definition makes little difference. To fairly observe the effect of a relay, we keep bursty user \mbox{traffic $p$} constant, and the new definition only expands or shrinks the two DoF regions by the same factor. Thus, we follow the convention in this work.


\section{Main Results}
\label{sec:mainresults}
For completeness, we first describe the following result for the single-user case, which is immediate since the cut-set bound is tight in terms of DoF in single-user networks~\cite{gamalkim:nit}.
\begin{theorem}
\label{thm:p2p}
The DoF of the bursty MIMO Gaussian relay channel is characterized by
\begin{align*}
d = \min \left\{ \begin{array}{l} p \min \left( M, N+L \right), p \min \left( M+L, N \right)
+ (1-p) \min \left( L, N \right)
\end{array} \right\}.
\end{align*}
\end{theorem}

Next, we present our main results for the bursty MIMO Gaussian interference channel (IC) with a relay.
\begin{theorem}
\label{thm:outer}
A DoF outer bound of the bursty MIMO Gaussian IC with a relay is 
\begin{align}
d_1, d_2 \leq & \min \left\{ \begin{array}{l}
p \min \left( M, N+L \right), 
p \min \left( M+L, N \right) + (1-p) \min \left( L, N \right)
\end{array} \right\}, \label{eq:ind} \\
d_1 + d_2 \leq & \min \left\{ \begin{array}{l}
p\min \left\{ (M-N)^+, N+L \right\} ,\\
p\min \left\{ (M+L-N)^+, N \right\} + (1-p)\min \left\{ (L-N)^+, N \right\}
\end{array} \right\} \notag \\
& + 
p^2 \min \left( 2M+L, N \right) + 2p(1-p) \min \left( M+L, N \right) + (1-p)^2\min \left( L, N \right) 
. \label{eq:sum}
\end{align}
\end{theorem}
\begin{IEEEproof}
See Section~\ref{sec:outerproof}.
\end{IEEEproof}

The above bound recovers the DoF results for the non-bursty case ($p=1$)~\cite{jafar:it09} and the case without a relay ($L=0$)~\cite{wang:spawc13}.

Using this bound, we obtain a necessary condition for attaining interference-free DoF. This is done by examining when \eqref{eq:sum} becomes inactive. The proof is in Appendix~\ref{app:necproof}.
\begin{corollary}
\label{cor:nec}
A necessary condition for attaining interference-free DoF is the union of three conditions $\mathcal{C}_1$, $\mathcal{C}_2$, and $\mathcal{C}_3$ below:
\begin{align*}
\mathcal{C}_1&: 2M \leq N, \\
\mathcal{C}_2&: M \geq 2N+L \text{ and } L \geq 2N, \\
\mathcal{C}_3&: M \geq 2N \text{ and } 3L \leq N.
\end{align*}
\end{corollary}

Finally, we establish a sufficient condition for attaining interference-free DoF.
\begin{theorem}
\label{thm:inner}
A sufficient condition for attaining interference-free DoF is the union of three conditions $\mathcal{C}_1,\mathcal{C}_2$ above, and $\mathcal{C}'_3$ below:
\begin{align*}
\mathcal{C}'_3: M \geq 2N+L \text{ and } 3L \leq N.
\end{align*}
\end{theorem}
\begin{IEEEproof}
See Section~\ref{sec:innerproof}.
\end{IEEEproof}

We briefly outline the schemes to be presented.
\begin{itemize}
\item $\mathcal{C}_1$: Each receiver can decode all symbols sent by the transmitters. The relay sends nothing at all times.
\item $\mathcal{C}_2$: All transmitting nodes can apply zero-forcing precoding. They can send symbols separately to each receiving node so that it does not get undesired symbols. But since the relay is shared by both users, it gets collided symbols when both transmitters are active. In our scheme, the relay {cooperates} with an active transmitter when it forwards them. This cooperation removes the interference in the collided symbols and delivers only the desired symbols to the receivers.
\item $\mathcal{C}'_3$: The transmitters can apply zero-forcing precoding, but the relay cannot. Since the relay cannot send symbols separately to each receiver, unavoidable collisions occur at the receivers. In our scheme, each transmitter provides the other receiver with {side information} so that the receivers exploit it to resolve the interference.
\end{itemize}

\begin{figure}[!t]
\centering
\includegraphics[width=0.5\columnwidth]{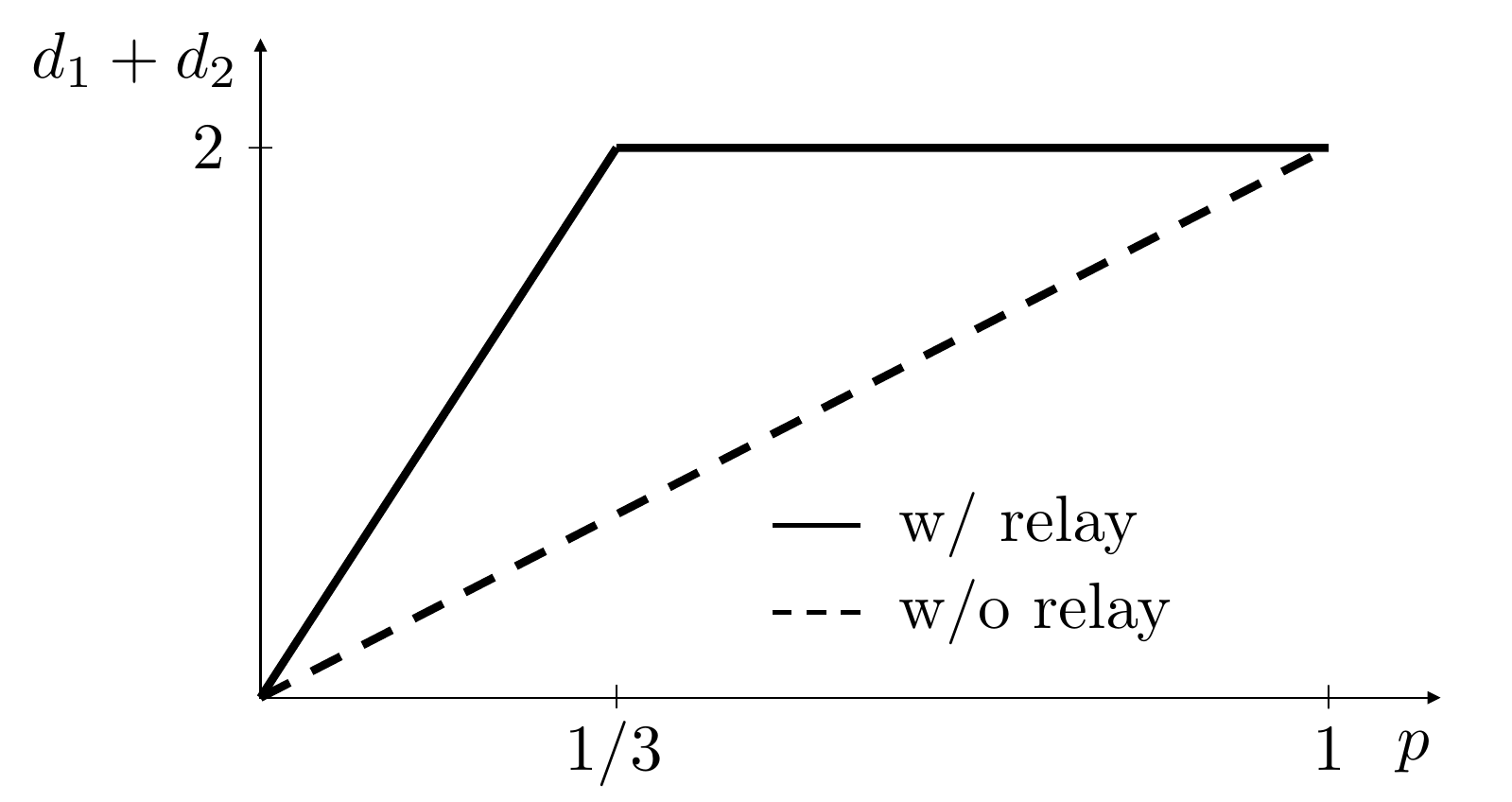}
\caption{Sum DoF of the bursty MIMO Gaussian IC with and without a relay. $(M,N,L)=(4,1,2)$ and $(M,N,L)=(4,1,0)$.}
\label{fig:sumdof}
\end{figure}

\begin{figure}[!t]
\centering
\includegraphics[width=0.5\columnwidth]{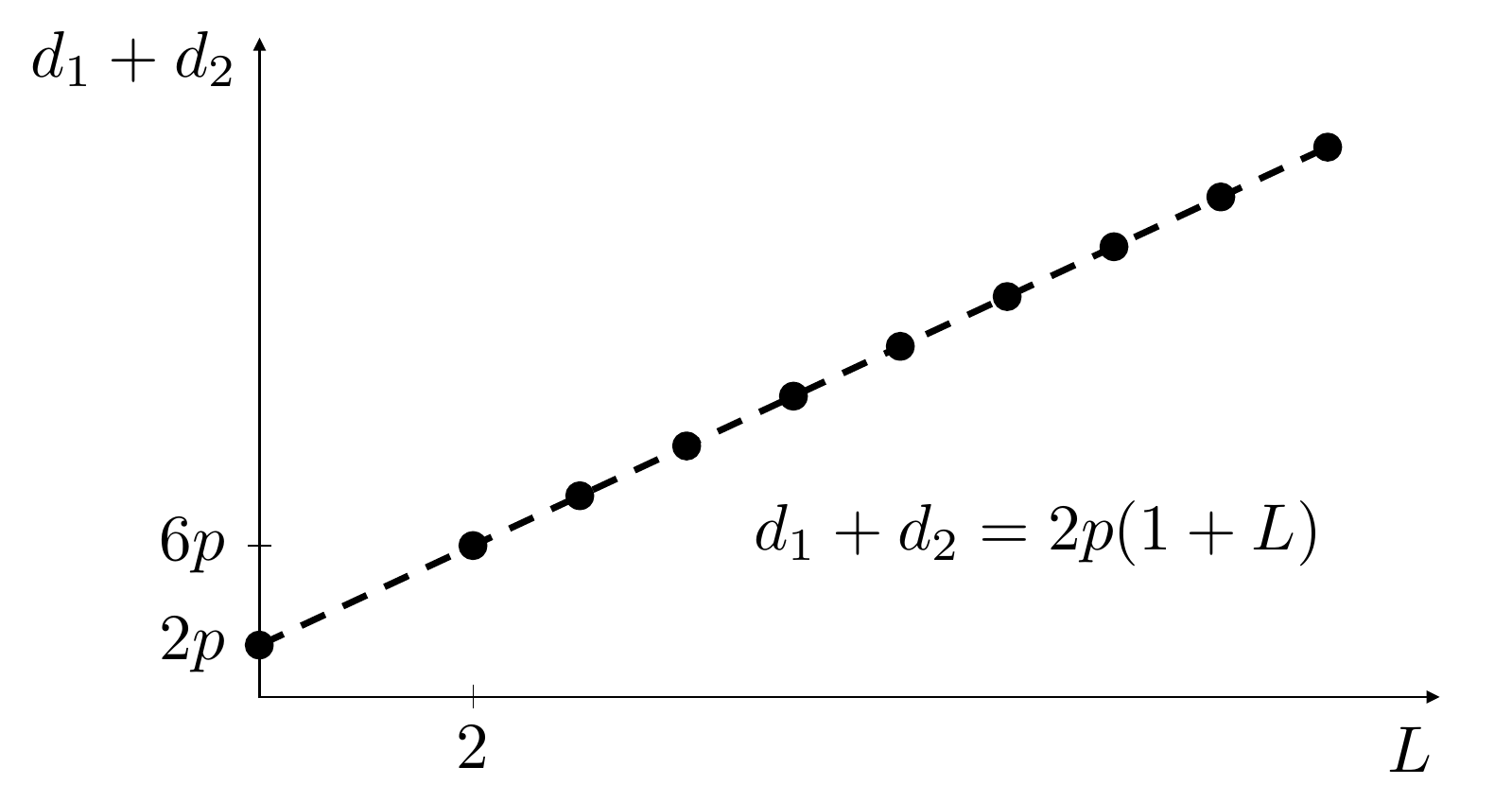}
\caption{Linear scalability of the sum DoF of the bursty MIMO Gaussian IC with a relay in low-traffic regimes $p \ll 1$. $M=\infty$, $N=1$, and $L \geq 2$.}
\label{fig:linear}
\end{figure}

Fig.~\ref{fig:sumdof} illustrates the sum DoF of the bursty MIMO Gaussian IC with and without a relay. We compare two antenna configurations $(M,N,L)=(4,1,2)$ and $(M,N,L)=(4,1,0)$. We can observe that the relay offers a DoF gain. Fig.~\ref{fig:linear} illustrates linear scalability of the sum DoF of the bursty MIMO Gaussian IC with a relay in low-traffic regimes $p \ll 1$. We consider a class of antenna configurations in which the relay has multiple antennas: $M=\infty$, $N=1$, and $L \geq 2$. We can observe that the sum DoF grows linearly with the number of antennas at the relay.

In this work, our results do not characterize the DoF region of the bursty MIMO Gaussian IC with a relay. We mainly discuss interference-free DoF performances and aim to show their optimality. However, we can establish the DoF region of the bursty SISO Gaussian IC with a multi-antenna relay.
\begin{theorem}
\label{thm:siso}
The DoF region of the bursty SISO Gaussian IC with a multi-antenna relay is
\begin{align*}
\mathcal{D} = \Big\{ (d_1, d_2) : d_1, d_2 \leq p, \ d_1 + d_2 \leq \min (2p, 1) \Big\}.
\end{align*}
\end{theorem}
\begin{IEEEproof}
See Appendix~\ref{app:siso}.
\end{IEEEproof}

\section{Proof of Theorem~\ref{thm:inner}}
\label{sec:innerproof}
In this section, we develop an explicit scheme that achieves interference-free DoF in the bursty MIMO Gaussian IC with a relay. We consider two different regimes depending on the level of data traffic.
\begin{itemize}
\item Low-traffic regime: Low data traffic limits the information flow at the transmitters. Thus, the transmitters send as much information as possible per active transmission toward the intended receivers and the relay.
\item High-traffic regime: High data traffic may limit the information flow at the receivers, especially when they have a small number of antennas. In this case, the transmitters reduce the amount of information sent per active transmission to ensure the decoding at the intended receivers.
\end{itemize}
\subsection{$2M \leq N$}
\label{subsec:120}
Each transmitter sends $M$ fresh symbols at all times, and the relay sends nothing. Since each receiver has a sufficient number of antennas, it decodes its desired symbols whenever its corresponding transmitter is active. This scheme achieves the following DoF region.
\begin{align*}
\mathcal{D} = \Big\{ (d_1, d_2) : d_1, d_2 \leq pM \Big\}.
\end{align*}

\subsection{$M \geq 2N+L$ and $L \geq 2N$}
\label{subsec:412}

\begin{figure*}[!t]
\centering
\includegraphics[width=\textwidth]{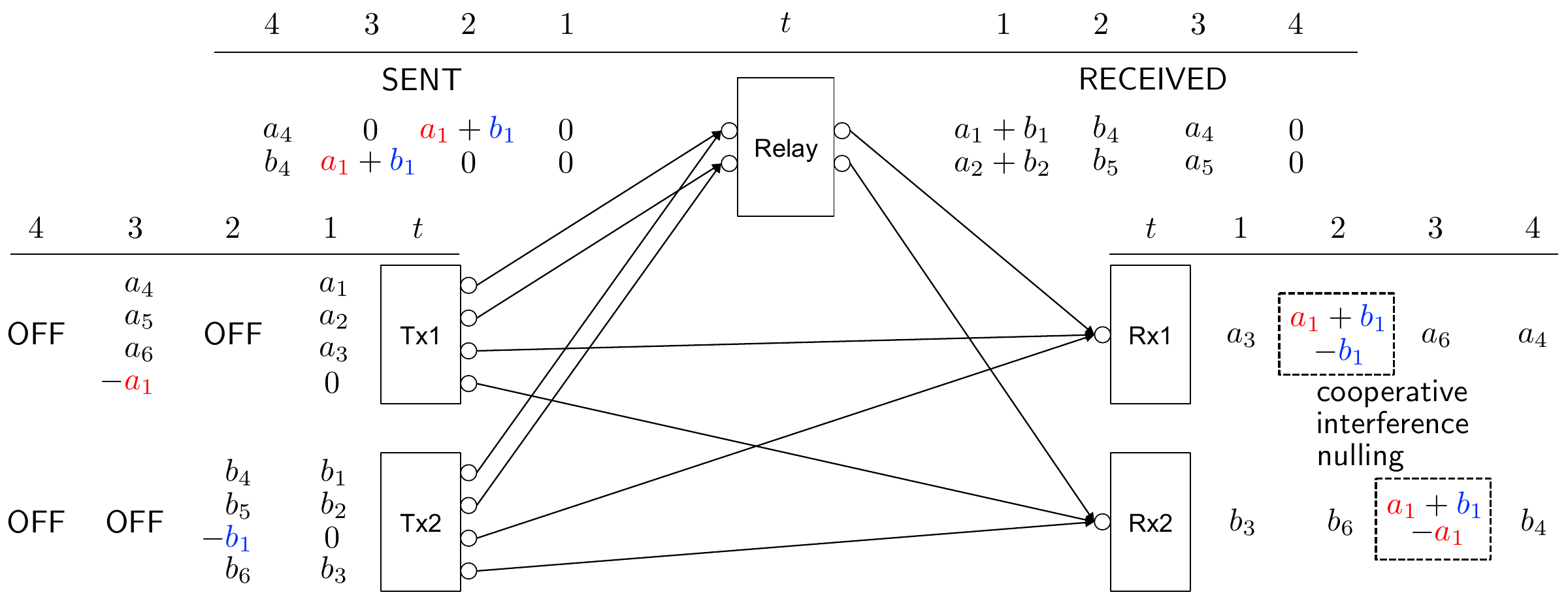}
\caption{An achievable scheme for ($M,N,L$) = ($4,1,2$) configuration.}
\label{fig:412achieve}
\end{figure*}

We present the scheme with an example for the simplest antenna configuration: ($M,N,L$) = (4,1,2). Fig.~\ref{fig:412achieve} demonstrates how the transmitters and the relay operate with a sequence of traffic states ($S_1,S_2$): (1,1), (0,1), (1,0), (0,0). {The generalization of the scheme is in Appendix~\ref{app:412gen} with a special case to note, which is a series of successive (1,1) traffic states.} 
The transmitters and the relay always apply zero-forcing precoding. The connected links in Fig.~\ref{fig:412achieve} depict the effect of zero-forcing precoding.

\textbf{Low-traffic Regime $p < \frac{1}{3}$:} Each transmitter sends one fresh symbol to its intended receiver and two to the relay at the rate of $p$. The use of one extra antenna is to send symbols to the other receiver when {cooperation} with the relay is needed.

\begin{itemize}
\item {Time 1:} Both transmitters send three fresh symbols. Knowing both are active through information of the current traffic states, the relay sends nothing to avoid interference at the receivers. However, two unavoidable collisions take place at the relay: $a_1+b_1$ and $a_2+b_2$.

\item {Time 2:} From feedback, the transmitters are aware of the past collisions at the relay. In addition to three fresh symbols toward \mbox{receiver 2} and the relay, \mbox{transmitter 2} sends $-b_1$ to \mbox{receiver 1} in hopes of removing its footprint in one of the past collisions, $a_1+b_1$. Knowing that \mbox{transmitter 2} is trying to deal with the collision, the relay sends $a_1+b_1$ to \mbox{receiver 1} and nothing to \mbox{receiver 2}. \mbox{Receiver 1} gets $a_1$, the sum of $a_1+b_1$ from the relay and $-b_1$ from \mbox{transmitter 2}. The relay and \mbox{transmitter 2} \emph{cooperate} and deliver desired symbol $a_1$ to \mbox{receiver 1} with interference $b_1$ removed.

\item {Time 3:} In addition to three fresh symbols, \mbox{transmitter 1} sends $-a_1$ to \mbox{receiver 2}. The relay sends $a_1+b_1$ to \mbox{receiver 2} and nothing to \mbox{receiver 1}. The cooperation between the relay and \mbox{transmitter 1} delivers $b_1$ to \mbox{receiver 2}.

\item {Time 4:} Both transmitters are inactive, and the relay knows there is no active transmitter to cooperate with. The relay delivers two past reserved symbols that were not collided. It sends $a_4$ to \mbox{receiver 1} and $b_4$ to \mbox{receiver 2}.
\end{itemize}

The proposed scheme works when relay-passing symbols, such as $a_1$, $b_1$, $a_4$, and $b_4$, are delivered to the intended receivers faster than they build up at the relay. Let us perform a simple analysis that compares the rate at which relay-passing symbols build up and the rate at which they are delivered. By the symmetry of the scheme, it suffices to carry out the analysis from \mbox{user 1}'s perspective. The analysis deals with two types of \mbox{user 1}'s relay-passing symbols. 
\begin{itemize}
\item Type 1 symbols: collisions ($a_1$, $a_2$). Two symbols, collided with symbols of the other user, are reserved at the relay with probability $p^2$ (\mbox{Time 1}: both transmitters are active.), and one of them can be delivered to \mbox{receiver 1} through \emph{cooperation} between the relay and \mbox{transmitter 2} with probability $(1-p)p$ (\mbox{Time 2}: only \mbox{transmitter 2} is active.). \mbox{Type 1} symbols are delivered faster than they build up at the relay when the following holds.
\begin{align*}
p^2 \times 2 < (1-p)p \times 1.
\end{align*}
\item Type 2 symbols: collision-free ($a_4$, $a_5$). Two symbols, without being collided, are reserved at the relay with probability $p(1-p)$ (\mbox{Time 3}: only \mbox{transmitter 1} is active.), and one of them can be delivered to \mbox{receiver 1} by the relay with probability $(1-p)^2$ (\mbox{Time 4}: both transmitters are inactive.). \mbox{Type 2} symbols are delivered faster than they build up at the relay when the following holds.
\begin{align*}
p(1-p) \times 2 < (1-p)^2 \times 1.
\end{align*}
\end{itemize}

In the low-traffic regime where $p < \frac{1}{3}$, the above conditions hold. Each transmitter sends \mbox{3 fresh} symbols at the rate of $p$, and all of them will be eventually decoded at the intended receiver: the individual DoF of $3p$.

\textbf{High-traffic Regime $\frac{1}{3} \leq p < 1$:} Both transmitters send symbols at a \emph{lower} rate; each transmitter chooses to send symbols with probability $q$ at any time instant. Each transmitter makes such decisions independently over time, and the decisions of the transmitters are independent. Therefore, each transmitter is in fact active with probability $pq$. A similar analysis by replacing $p$ with $pq$ gives the following conditions.
\begin{itemize}
\item Type 1 symbols: $(pq)^2 \times 2 < (1-pq)(pq) \times 1$.
\item Type 2 symbols: $(pq)(1-pq) \times 2 < (1-pq)^2 \times 1$.
\end{itemize}

In the high-traffic regime where $\frac{1}{3} \leq p < 1$, defining $q$ as $\frac{1}{p}(\frac{1}{3} - \epsilon)$, where $\epsilon > 0$, satisfies the above conditions. Each transmitter sends \mbox{3 fresh} symbols at the rate of $pq$, and all of them will be eventually decoded at the intended receiver: the individual DoF of $3pq$. As both transmitters choose $\epsilon$ arbitrarily close to zero, the individual DoF converges to 1. In summary, the proposed scheme achieves the following DoF region.\begin{align*}
\mathcal{D} = \Big\{ (d_1, d_2) : d_1, d_2 \leq \min ( 3p, 1 ) \Big\}.
\end{align*}

\begin{remark}[Cooperative Interference Nulling]
From \mbox{user 1}'s perspective, to achieve interference-free DoF performances, \mbox{transmitter 1} should always send three fresh symbols as in the single-user case. One of them is directly delivered to \mbox{receiver 1}, and the other two are reserved at the relay and delivered later when \mbox{transmitter 1} is inactive. Unfortunately, since the relay is shared, the relay-passing symbols of \mbox{user 1} sometimes get interfered with those of \mbox{user 2}. But, cooperative interference nulling removes the interference in the \mbox{user 1}'s relay-passing symbols and delivers only the desired symbols to \mbox{receiver 1}. At \mbox{Time 2}, for example, when \mbox{transmitter 1} is inactive, the relay and active \mbox{transmitter 2} cooperate and remove \mbox{interference $b_1$} in $a_1+b_1$ to deliver desired \mbox{symbol $a_1$} to \mbox{receiver 1}. When both transmitters are inactive, the relay applies zero-forcing precoding and delivers \mbox{user 1}'s relay-passing symbols that were not interfered, for example, $a_4$ at \mbox{Time 4}. Overall, the operation coincides with the single-user scheme: \mbox{transmitter 1} always sends three symbols. One of them is directly delivered to \mbox{receiver 1}, and the other two are delivered through the relay to \mbox{receiver 1} without interference.
\end{remark}

\begin{remark}[Distinction from Other Relaying Schemes]
Cooperative interference nulling is a notable distinction from other relaying schemes. It provides a DoF gain in the bursty IC, although other relaying schemes provide only power gains in the non-bursty IC. In cooperative interference nulling, the relay and the transmitters \emph{synchrounously} cooperate by exploiting information of the bursty traffic states, to remove interference in their signals in the air and deliver only desired signals to the receivers. In other schemes based on decode-forward strategies~\cite{sahin:07, goldsmith:it12}, the relay and the transmitters also cooperate by generating their signals coherently. However, the schemes provide only power gains. In other schemes mostly based on compress-forward strategies~\cite{yu:ita10, yener:it11, noisy:it11, kang:isit13, tobias:it14}, the relay forwards additional descriptions of its received signals to help the decoding without cooperating with the transmitters. Also, the schemes provide only power gains.

\end{remark}

\begin{remark}[Why There Is a DoF Gain in the Bursty Case, but None in the Non-Bursty Case] 
In the bursty case, transmitters sometimes become idle. In the idle moments, the presence of a relay is valuable as it can be a temporary information source for the receivers. Based on its past received symbols, the relay can either send symbols that resolve past collisions at the receivers, or send fresh symbols that were reserved at the relay and have not been delivered to the receivers. In both ways, the relay enables a receiver or both to decode extra symbols while a transmitter or both are idle. These extra symbols amount to a DoF gain. In the non-bursty case, however, transmitters never become idle. It means that whatever symbols the relay can possibly deliver to the receivers based on its past received symbols, the transmitters can deliver the same symbols to the receivers by themselves: be it symbols that resolve past collisions, or fresh symbols. The relay finds no moments to be as valuable as it can be in the bursty case where transmitters sometimes send no useful symbols at all being idle.
\end{remark}

\subsection{$M \geq 2N+L$ and $3L \leq N$}
\label{subsec:731}

\begin{figure*}[!t]
\centering
\includegraphics[width=\textwidth]{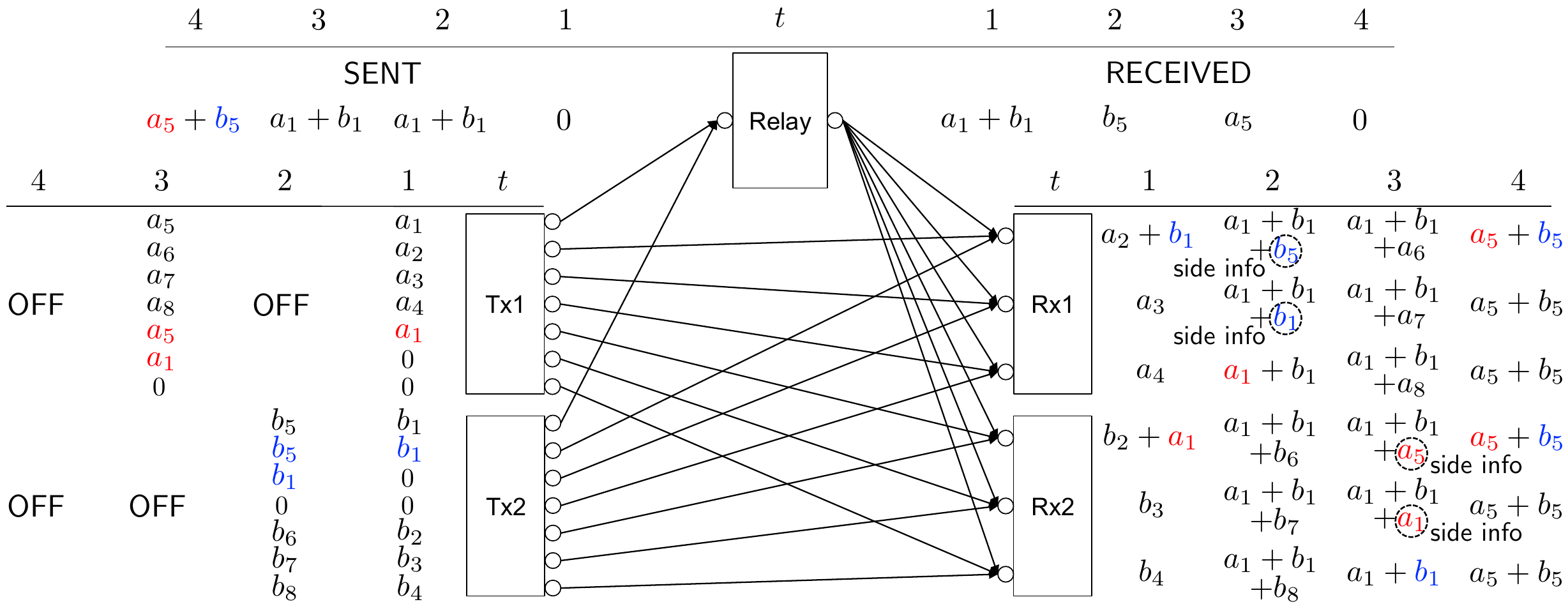}
\caption{An achievable scheme for ($M,N,L$) = (7,3,1) configuration.}
\label{fig:731achieve}
\end{figure*} 

We present the scheme with an example for the simplest antenna configuration: ($M,N,L$) = (7,3,1). Fig.~\ref{fig:731achieve} demonstrates how the transmitters and the relay operate with a sequence of traffic states ($S_1,S_2$): (1,1), (0,1), (1,0), (0,0). {The generalization of the scheme is in Appendix~\ref{app:731gen}.} 
The transmitters always apply zero-forcing precoding, whereas the relay cannot. The connected links in Fig.~\ref{fig:731achieve} depict the effect of zero-forcing precoding.

\textbf{Low-traffic Regime $p < \frac{1}{2}$:} Each transmitter sends three fresh symbols to its intended receiver and one to the relay at the rate of $p$. The use of three extra antennas is to provide the other receiver with {side information} that is needed to resolve unavoidable collisions.

\begin{itemize}
\item {Time 1:} Both transmitters send four fresh symbols. In addition, each transmitter sends to the other receiver the duplicate of its relay-passing symbol. This is to provide side information.
There is one collision at the relay, $a_1 + b_1$, and one at each receiver, $a_2+b_1$ and $b_2+a_1$. For each receiver, this collision has to be resolved to decode its desired symbol, $a_2$ and $b_2$.

\item {Time 2:} In addition to four fresh symbols, \mbox{transmitter 2} sends $b_1$ and $b_5$, the duplicate of its relay-passing symbols, to \mbox{receiver 1}. This is again to provide {side information}.
The relay broadcasts $a_1 + b_1$ to deliver $a_1$ to \mbox{receiver 1} whose corresponding transmitter is inactive. \mbox{Receiver 1} decodes $a_1$, $b_1$, and $b_5$. \mbox{Receiver 1} resolves past collision $a_2+b_1$ with \emph{side information} $b_1$. $b_5$ will be used later.

\item {Time 3:} In addition to four fresh symbols, \mbox{transmitter 1} sends $a_1$ and $a_5$ to \mbox{receiver 2}. The relay broadcasts $a_1+b_1$. \mbox{Receiver 2} decodes $b_1$, $a_1$, and $a_5$. \mbox{Receiver 2} resolves past collision $b_2 + a_1$ with side information $a_1$. $a_5$ will be used later.

\item {Time 4:} Both transmitters are inactive. The relay sends the {sum} of $a_5$ and $b_5$ to deliver information that is useful for {both} receivers. From $a_5+b_5$, \mbox{receiver 1} decodes $a_5$ since it has $b_5$ as side information, and \mbox{receiver 2} decodes $b_5$ since it has $a_5$ as side information.
\end{itemize}

The proposed scheme works when relay-passing symbols, such as $a_1$, $b_1$, $a_5$, and $b_5$, are delivered to the intended  receivers faster than they build up at the relay. Also, each receiver needs all relay-passing symbols of the other user as side information, because they are broadcast by the relay and cause interference. This is why each transmitter keeps trying to provide the other receiver with side information at the cost of unnecessary interference, for example, at \mbox{Time 1}. Let us perform a simple analysis. First, we compare the rate at which relay-passing symbols build up and the rate at which they are delivered. Second, we examine the rate at which side information is provided. Due to the symmetry of the scheme, it suffices to carry out the analysis from \mbox{user 1}'s perspective.
\begin{itemize}
\item User 1's relay-passing symbols ($a_1$, $a_5$): one symbol, possibly collided with a symbol of the other user, reserved at the relay with probability $p$ (\mbox{Time 1, 3}: \mbox{transmitter 1} is active.), and it can be delivered to \mbox{receiver 1} by the relay with probability $1-p$ (\mbox{Time 2, 4}: \mbox{transmitter 1} is inactive.). Transmitted symbols of the relay can be in the form of sums of both users' relay-passing symbols. Yet, \mbox{receiver 1} can decode \mbox{user 1}'s relay-passing symbols since \mbox{transmitter 2} provides side information properly. This matter is discussed in the next item. Considering \mbox{user 1}'s relay-passing symbols, they are delivered faster than they build up at the relay when the following holds.
\begin{align*}
p \times 1 < (1-p) \times 1.
\end{align*}

\item User 2's relay-passing symbols ($b_1$, $b_5$): one symbol is reserved at the relay with probability $p$ (\mbox{Time 1, 2}: \mbox{transmitter 2} is active.), and eventually broadcast. This causes interference at \mbox{receiver 1}. \mbox{Receiver 1} can get the duplicate of at most two \mbox{user 2}'s relay-passing symbols as \emph{side information} from \mbox{transmitter 2} with probability $(1-p)p$ (\mbox{Time 2}: only \mbox{transmitter 2} is active.). The reason why \mbox{receiver 1} gets at most two of them is that it uses one antenna to get one \mbox{user 1}'s relay-passing symbol from the relay, and has two antennas left. \mbox{Receiver 1} exploits this side information to resolve the interference caused by the broadcasting of the relay. Side information from \mbox{transmitter 2} is provided at a faster rate than the rate at which \mbox{user 2}'s relay-passing symbols build up at the relay when the following holds.
\begin{align*}
p \times 1 < (1-p)p \times 2.
\end{align*}
\end{itemize}

In the low-traffic regime where $p < \frac{1}{2}$, the above conditions hold. Each transmitter sends \mbox{4 fresh} symbols at the rate of $p$, and all of them will be eventually decoded at the intended receiver in the low-traffic regime: the individual DoF of $4p$.

\textbf{High-traffic Regime $\frac{1}{2} \leq p < 1$:} Both transmitters send fresh symbols to the relay at a \emph{lower} rate; they choose to send symbols to the relay with probability $q$ at any time instant. Each transmitter makes such decisions independently over time, and the decisions of the transmitters are independent. Therefore, each transmitter sends symbols to its intended receiver at the rate of $p$, and to the relay at the rate of $pq$. We can perform a similar analysis from \mbox{user 1}'s perspective.
\begin{itemize}
\item User 1's relay-passing symbols are reserved at the rate of $pq \times 1$, and they can be delivered at the rate of $(1-p) \times 1$.
\begin{align*}
pq \times 1 < (1-p) \times 1.
\end{align*}

\item User 2's relay-passing symbols are reserved at the rate of $pq \times 1$, and eventually broadcast. \mbox{Receiver 1} can get {side information} at the rate of $(1-p)p \times 2$.
\begin{align*}
pq \times 1 < (1-p)p \times 2.
\end{align*}
\end{itemize}

In the high-traffic regime where $ \frac{1}{2} \leq p < 1$, defining $q$ as $\frac{1}{p}(1-p- \epsilon)$, where $\epsilon > 0$, satisfies the above conditions. Each transmitter sends \mbox{3 fresh} symbols to its intended receiver at the rate of $p$, and \mbox{1 fresh} symbol to the relay at the rate of $pq$. All of them will be eventually decoded at the intended receiver: the individual DoF of $3p + 1pq$. As both transmitters choose $\epsilon$ arbitrarily close to zero, the individual DoF converges to $2p + 1$. In summary, the proposed scheme achieves the following DoF region.
\begin{align*}
\mathcal{D} = \Big\{ (d_1, d_2) : d_1, d_2 \leq \min ( 4p, 2p+1 ) \Big\}.
\end{align*}

\begin{remark}[Exploiting Side Information]
The limited number of antennas at the relay disables cooperative interference nulling. Moreover, transmitted symbols of the relay are broadcast to both receivers, thus each receiver unavoidably gets undesired relay-passing symbols of the other user. Each transmitter provides the other receiver with side information that resolves the unavoidable collisions. At \mbox{Time 2}, for example, active \mbox{transmitter 2} provides \mbox{receiver 1} with $b_1$ and $b_5$ as side information when \mbox{transmitter 1} is inactive, and \mbox{receiver 1} exploits them to resolve the interference in $a_2 + b_1$ and $a_5 + b_5$. Overall, \mbox{transmitter 1} always sends four fresh symbols to achieve interference-free DoF, and the symbols arrive at \mbox{receiver 1}, either directly or through the relay, sometimes interfered by relay-passing symbols of \mbox{user 2}. \mbox{Receiver 1} resolves the interference and decodes its desired symbols with side information provided by \mbox{transmitter 2}.
\end{remark}

\begin{remark}[Other Examples That Exploit Side Information]
The gain obtained by exploiting side information appears in many other network examples. In~\cite{Katti:ACM}, a wireless router receives multiple packets intended for different destinations from its neighboring nodes, encodes them into one packet, and broadcasts it. In a single transmission, every destination decodes its desired packet by using the packets for the other destinations that it overheard as side information. In~\cite{SuhTse}, channel output feedback can increase the non-feedback capacity of the Gaussian IC. In~\cite{MaddahAli:it12}, outdated channel state feedback can increase the non-feedback capacity of the Gaussian MIMO broadcast channel. In both works, feedback enables receivers to exploit their past received signals as side information.
\end{remark}

\section{Proof of Theorem~\ref{thm:outer}}
\label{sec:outerproof}
The bound~\eqref{eq:ind} is the cut-set bound, so we omit the proof. The bound~\eqref{eq:sum} consists of two bounds on $d_1 + d_2$, and we derive them in this section. In the following derivations, we can get two additional bounds on $R_1 + R_2$ by changing the order of $R_1$ and $R_2$. But these bounds on $R_1 + R_2$ result in the identical bounds on $d_1 + d_2$ due to symmetry, so we omit the proofs of them. The outer bound proof follows the genie-aided approach. For notational convenience, let $\sum$ denote $\sum_{t=1}^{n}$, $S_t$ denote $(S_{1t}, S_{2t})$, and $S^n$ denote the sequence of $S$ up to $n$.

One bound on $R_1+R_2$ can be derived as follows.
\begin{align*}
& n(R_1 + R_2 - \epsilon_n) \overset{(a)}{\leq} I(W_1;Y_1^n, S^n) + I(W_2;Y_2^n, S^n) \overset{(b)}{\leq} I(W_1;Y_1^n| S^n) + I(W_2;Y_1^n, Y_2^n, Y_R^n| S^n, W_1) \\
& \overset{(c)}{\leq} \sum h(Y_{1t}|S^n, Y_1^{t-1}) - \sum h(Y_{1t}| S^n, W_1, W_2, Y_1^{t-1}, Y_2^{t-1}, Y_R^{t-1}) \\
& \ \quad +  \sum h(Y_{2t}, Y_{Rt} | S^n, W_1, Y_1^{t-1}, Y_2^{t-1}, Y_R^{t-1}, Y_{1t}) - \sum h(Y_{2t}, Y_{Rt} | S^n, W_1, W_2, Y_1^{t-1}, Y_2^{t-1}, Y_R^{t-1}, Y_{1t}) \\
& \overset{(d)}{\leq} \sum h(Y_{1t}|S_t) - \sum h(Z_{1t})  + \sum h(Y_{2t}, Y_{Rt} | S_t, X_{1t}, X_{Rt}, Y_{1t}) - \sum h(Z_{2t}, Z_{Rt}) \\
& \overset{(e)}{\leq} p^2 \sum h \left( \mathbf{H}_{11}{X}_{1t} + \mathbf{H}_{12}{X}_{2t} + \mathbf{H}_{1R}X_{Rt} + Z_{1t} \right) + p(1-p) \sum h \left( \mathbf{H}_{11}{X}_{1t} + \mathbf{H}_{1R}X_{Rt} + Z_{1t} \right) \\
& \ \quad + (1-p)p \sum h \left( \mathbf{H}_{12}{X}_{2t} + \mathbf{H}_{1R}X_{Rt} + Z_{1t} \right) + (1-p)^2 \sum h \left( \mathbf{H}_{1R}X_{Rt} + Z_{1t} \right) - \sum h \left( Z_{1t} \right) \\
& \ \quad + p \sum h \left( \mathbf{H}_{22}{X}_{2t}+Z_{2t}, \mathbf{H}_{R2}{X}_{2t}+Z_{Rt} \middle| \mathbf{H}_{12}{X}_{2t} + Z_{1t} \right) - p \sum h \left( Z_{2t}, Z_{Rt} \right),
\end{align*}
where $(a)$ is from Fano's inequality; 
$(b)$ is from the mutual independence of ($W_1$, $W_2$, $S^n$); $(c)$ is from conditioning reduces entropy; $(d)$ is from $X_{kt} = f_{kt}(W_k, S_{kt}, S^{t-1})$ and $X_{Rt} = f_{Rt}(Y_R^{t-1}, S^t)$, the mutual independence of ($Z_1^n$, $Z_2^n$, $Z_R^n$, $W_1$, $W_2$, $S^n$), the i.i.d. assumption of ($Z_1^n$, $Z_2^n$, $Z_R^n$), and conditioning reduces entropy; $(e)$ is from conditioning reduces entropy, and the evaluation of $S_t$.

The other bound on $R_1+R_2$ can be derived as follows.
\begin{align*}
& n(R_1 + R_2 - \epsilon_n) \overset{(a)}{\leq} I(W_1;Y_1^n, S^n) + I(W_2;Y_2^n, S^n) \overset{(b)}{\leq} I(W_1;Y_1^n| S^n) + I(W_2;Y_1^n, Y_2^n| S^n, W_1) \\
& = \sum I(W_{1};Y_{1t}|S^n, Y_1^{t-1}) + \sum I(W_2;Y_{1t}| S^n, W_1, Y_1^{t-1}, Y_2^{t-1}) + \sum I(W_2;Y_{2t} | S^n, W_1, Y_1^{t-1}, Y_2^{t-1}, Y_{1t}) \\
& \overset{(c)}{\leq} \sum h(Y_{1t}|S^n, Y_1^{t-1}) - \sum h(Y_{1t}| S^n, W_1, W_2, Y_1^{t-1}, Y_2^{t-1}, Y_R^{t-1}) \\
& \ \quad + \sum h(Y_{2t}| S^n, W_1, Y_1^{t-1}, Y_2^{t-1}, Y_{1t}) - \sum h(Y_{2t}| S^n, W_1, W_2, Y_1^{t-1}, Y_2^{t-1}, Y_R^{t-1}, Y_{1t}) \\
& \overset{(d)}{\leq} \sum h(Y_{1t}|S_t) - \sum h(Z_{1t}) + \sum h(Y_{2t} | S_t, X_{1t}, Y_{1t}) - \sum h(Z_{2t}) \\
& \overset{(e)}{\leq} p^2 \sum h \left( \mathbf{H}_{11}{X}_{1t} + \mathbf{H}_{12}{X}_{2t} + \mathbf{H}_{1R}X_{Rt} + Z_{1t} \right) + p(1-p) \sum h \left( \mathbf{H}_{11}{X}_{1t} + \mathbf{H}_{1R}X_{Rt} + Z_{1t} \right) \\
& \ \quad + (1-p)p \sum h \left( \mathbf{H}_{12}{X}_{2t} + \mathbf{H}_{1R}X_{Rt} + Z_{1t} \right) + (1-p)^2 \sum h \left( \mathbf{H}_{1R}X_{Rt} + Z_{1t} \right) - \sum h \left( Z_{1t} \right) \\
& \ \quad + p \sum h \left( \mathbf{H}_{22}{X}_{2t} + \mathbf{H}_{2R}{X}_{Rt} + Z_{2t} \middle|  \mathbf{H}_{12}{X}_{2t} + \mathbf{H}_{1R}{X}_{Rt} + Z_{1t} \right) \\
& \ \quad + (1-p) \sum h \left( \mathbf{H}_{2R}{X}_{Rt} + Z_{2t} \middle| \mathbf{H}_{1R}{X}_{Rt} + Z_{1t} \right)  - \sum h \left( Z_{2t} \right),
\end{align*}
where $(a)$ is from Fano's inequality; $(b)$ is from the mutual independence of ($W_1$, $W_2$, $S^n$); $(c)$ is from conditioning reduces entropy; $(d)$ is from $X_{kt} = f_{kt}(W_k, S_{kt}, S^{t-1})$ and $X_{Rt} = f_{Rt}(Y_R^{t-1}, S^t)$, the mutual independence of ($Z_1^n$, $Z_2^n$, $Z_R^n$, $W_1$, $W_2$, $S^n$), the i.i.d. assumption of ($Z_1^n$, $Z_2^n$), and conditioning reduces entropy; $(e)$ is from conditioning reduces entropy, and the evaluation of $S_t$.

To get the claimed outer bound on $d_1 + d_2$, we evaluate the above bounds with the Gaussian distributions that maximize the differential entropies~\cite{gamalkim:nit}, and take the limit as $P \rightarrow \infty$ after dividing them by $\log(P)$.

\section{Discussion}
\label{sec:discussion}
\textit{On Optimality of Our Schemes:} In this work, we established a necessary condition and a sufficient condition for interference-free DoF performances, but they are not identical. For a class of antenna configurations ($\mathcal{C}_3-\mathcal{C}'_3$), optimality in terms of DoF was not shown. For a resolved class ($\mathcal{C}_1$), a simple and naive scheme achieves optimality. Both transmitters send signals and the relay sends nothing. For the other resolved classes ($\mathcal{C}_2$ and $\mathcal{C}'_3$), our scheme relies on zero-forcing precoding of the transmitters, through which the transmitters minimize the reception of undesired symbols at the receiving nodes and resolve unavoidable collisions in cooperation with the relay. For the unresolved class ($\mathcal{C}_3-\mathcal{C}'_3$), however, it is straightforward to see that each transmitter can no longer apply complete zero-forcing precoding, thus it loses control over the destinations of its transmitted symbols. It becomes more difficult for the transmitters not only to avoid undesirable collisions at the receiving nodes, but also to cooperate with the relay. It is still not clear if we need a better achievable scheme, a tighter outer bound, or both.

\textit{Model Assumptions:} In formulating the model in question, we assumed independence between the two traffic states, considering the case where the transmitters lack coordination in sharing a common communication medium. However, it may be too pessimistic to discard the idea of well designed distributed media access control protocols that can enable good coordination in wireless systems. Thus, we can consider a model in which the transmitters can instantaneously cooperate. Then, \mbox{transmitter $k$} generates its encoded signal at \mbox{time $t$} based on its own message and the traffic states of both transmitters up to $t$: $X_{kt} =  f_{kt}(W_k, S^t)$, $\forall k = 1,2$. Also, it may be interesting to consider a non-causal model in which each tranasmitter knows all its own traffic states including its future traffic states, and generates its encoded signals accordingly: $X_{kt} = f_{kt}(W_k, S_k^n, S_l^{t-1})$, where $l$ refers to the other transmitter. In both models, our outer bound proof in Section~\ref{sec:outerproof} is still valid. Thus, in the models where the transmitters instantaneously cooperate and where each transmitter organizes its transmissions in advance knowing its own future traffic states, our necessary condition for interference-free DoF performances remains valid. For the resolved classes of antenna configurations ($\mathcal{C}_1$, $\mathcal{C}_2$, and $\mathcal{C}'_3$), our schemes, which assume neither instantaneous coordination between the transmitters nor a priori individual coordination of transmissions, can achieve optimality in terms of DoF. However, it is unclear whether or not the DoF regions for the aforementioned unresolved class of antenna configurations ($\mathcal{C}_3-\mathcal{C}'_3$) are also identical.

\textit{Future Works:} One can explore the complete coordination model or the non-causal model to narrow the gap between our necessary condition and our sufficient condition for interference-free DoF performances. Gained insights from these two models may then lead to showing optimality in the model of our work. Also, one can further look into the role of relays in $K$-user bursty interference networks to see if they provide significant gains, such as interference-free performances with increased DoF and scalability of DoF gains.

\section{Conclusion}
\label{sec:conclusion}
We discovered that an in-band relay can provide a DoF gain in the two-user {bursty} MIMO Gaussian IC. We demonstrated that the relay can help achieve {interference-free} DoF performances for certain antenna configurations. The relay and the transmitters cooperate by exploiting information of the bursty traffic states to achieve the performances. Moreover, we observed that the gain can be particularly substantial with low data traffic, as it can grow {linearly} with the number of antennas at the relay. Our results show promising benefits that relays can offer in practical communication systems where multiple source-destination links interfere with each other in a bursty manner due to intermittent data traffic.


%

\appendices

\section{Proof of Corollary~\ref{cor:nec}}
\label{app:necproof}
This appendix proves the necessary condition for attaining interference-free DoF in Corollary~\ref{cor:nec} by examining when \eqref{eq:sum} becomes inactive. 

\subsection{$2M \leq N$}
{Individual DoF bound:} $M < N+L$ gives us
\begin{align*}
d_1, d_2 & \leq \min \left\{
p(M), p\min ( M+L, N ) + (1-p)\min ( L, N )
\right\}.
\end{align*}

From $M \leq M+L$ and $M < N$, we have $p(M) \leq p \min ( M+L, N )$. Hence, the individual DoF is bounded by
\begin{align*}
d_1, d_2 \leq pM.
\end{align*}

{Sum DoF bound:} $ M < N $ gives us
\begin{align*}
d_1 + d_2 & \leq 
p^2 \min ( 2M+L, N ) + 2p(1-p) \min ( M+L, N ) + (1-p)^2 \min ( L, N ).
\end{align*}

From $ 2M \leq 2M+L $ and $2M \leq N$, we have $p^2 (2M) \leq p^2 \min ( 2M+L, N )$. From $M \leq M+L$ and $M<N$, we have $ 2p(1-p)(M) \leq 2p(1-p) \min (M+L, N)$. The sum of these two inequalities shows that the sum of the two individual DoF bounds is tighter than the sum DoF bound. Hence, the sum DoF is bounded by
\begin{align*}
d_1 + d_2 \leq 2pM.
\end{align*}

Therefore, $2M \leq N$ is a necessary condition for attaining interference-free DoF for all $p<1$.

\subsection{$M \leq N < 2M$}
{Individual DoF bound:} $ M \leq N+L $ gives us
\begin{align*}
d_1, d_2 & \leq \min \left\{
p(M), p\min ( M+L, N ) + (1-p) \min ( L, N )
\right\}.
\end{align*}

From $M \leq M+L$ and $M \leq N$, we have $ p(M) \leq p\min ( M+L, N )$. Hence, the individual DoF is bounded by
\begin{align*}
d_1, d_2 \leq pM.
\end{align*}

{Sum DoF bound:} $ M \leq N $ and $2M + L > N$ give us
\begin{align*}
d_1 + d_2 & \leq 
p^2 (N) + 2p(1-p) \min ( M+L, N ) + (1-p)^2 \min ( L, N ).
\end{align*}

If we choose a looser sum DoF bound and show the looser sum DoF bound is strictly tighter than the sum of the two individual DoF bounds, then we show the actual sum DoF bound is also strictly tighter. From $ \min ( M+L, N ) \leq N $ and $ \min (L, N) \leq N $, we have $ 2p(1-p) \min ( M+L, N ) \leq 2p(1-p) (N) $ and $ (1-p)^2 \min (L, N) \leq (1-p)^2 (N) $. Hence, the sum DoF is bounded by
\begin{align*}
d_1 + d_2 \leq N.
\end{align*}

For $\frac{N}{2M} < p < 1$, the looser sum DoF bound is strictly tighter than the sum of the two individual DoF bounds. 

Therefore, $ M \leq N < 2M $ is not a necessary condition for attaining interference-free DoF for all $ p < 1$.

\subsection{$N < M < 2N$}
{Individual DoF bound:} $M + L > N$ gives us 
\begin{align*}
d_1,d_2 \leq \min \left\{
p \min ( M, N+L ), p (N) + (1-p) \min ( L, N )
\right\}.
\end{align*}

{Sum DoF bound:} $M - N < N + L$, $2M+L > N$, and $M+L > N$ give us
\begin{align*}
d_1 + d_2 & \leq \min \left\{
p (M-N), p \min ( M+L-N, N ) + (1-p) \min \{ (L-N)^+, N \}
\right\} \\
& + p^2 (N) + 2p(1-p) (N) + (1-p)^2 \min ( L, N )
.
\end{align*}

From $M-N \leq M+L-N$ and $M-N < N$, we have $p(M-N) \leq p \min ( M+L-N, N )$. Hence, the sum DoF is bounded by
\begin{align*}
d_1 + d_2 & \leq p (M-N) + 
p^2 (N) + 2p(1-p) (N) + (1-p)^2 \min ( L, N )
.
\end{align*}

Given $M$, $N$, and $L$, the difference between the two terms of the big minimum function in the individual DoF bound is a continuous function $f(p)$ in $p \in [0,1]$. When $L>0$, by the intermediate value theorem, there exists $p_0 \in [0,1]$ such that $f(p_0) = 0$ since $f(0)f(1) < 0$. And, $p_0 \in (0,1)$ since $f(0) \neq 0$ and $f(1) \neq 0$. Hence, the second term is active for $p_0 \leq p < 1$. When $L=0$, the second term is always active.

For $ p_0 \leq p < 1$, the following inequality should necessarily hold to attain interference-free DoF.
\begin{align*}
2p(N) + 2(1-p)\min ( L, N ) \leq p (M-N) + 
p^2 (N) + 2p(1-p) (N) + (1-p)^2 \min ( L, N ).
\end{align*}

When $L \geq N$, the inequality becomes
\begin{align*}
p \geq \frac{N}{M-N}.
\end{align*}
The above inequality does not hold for $ p_0 \leq p <1$.

When $L < N$, the inequality becomes
\begin{align*}
p^2(L-N) + p(M-N) - L \geq 0.
\end{align*}

From $ L < N $, $g(p) = p^2(L-N) + p(M-N) - L$ is a strictly concave quadratic function. $g(1) \geq 0$ should necessarily hold for the above inequality to hold for $ p_0 \leq p < 1$. But $g(1) = M - 2N < 0$.

Therefore, $N < M < 2N$ is not a necessary condition for attaining interference-free DoF for all $p<1$.

\subsection{$M \geq 2N$ and $L < N$}
{Individual DoF bound:} $M > N+L$, $ M+L > N $, and $L < N$ give us 
\begin{align*}
d_1, d_2 & \leq \min \left\{
p(N+L), p(N) + (1-p)(L) 
\right\}.
\end{align*}

{Sum DoF bound:} $ M+L-N \geq N $, $L < N$, $ 2M + L > N $, and $  M + L > N$ give us
\begin{align*}
d_1 + d_2 & \leq \min \left\{
p \min ( M-N, N+L ), p (N)
\right\} + p^2 (N) + 2p(1-p) (N) + (1-p)^2 (L).
\end{align*}

From $M-N \geq N$ and $N+L \geq N$, we have $p \min ( M-N, N+L ) \geq p(N)$. Hence, the sum DoF is bounded by
\begin{align*}
d_1 + d_2 & \leq p(N) + p^2 (N) + 2p(1-p) (N) + (1-p)^2 (L).
\end{align*}

For $p < \frac{1}{2}$, in the individual DoF bound, $p(N+L)$ term is active. Otherwise, $p(N)+(1-p)(L)$ term is active. 

For $p < \frac{1}{2}$, the following inequality should necessarily hold to attain interference-free DoF.
\begin{align*}
2p(N+L) \leq p(N) + p^2 (N) + 2p(1-p) (N) + (1-p)^2 (L).
\end{align*}
The inequality becomes $ p(1-p)N \geq (-p^2 + 4p - 1) L $.

For $\frac{1}{2} \leq p < 1$, the following inequality should necessarily hold to attain interference-free DoF.
\begin{align*}
2p(N)+2(1-p)(L) \leq p(N) + p^2 (N) + 2p(1-p) (N) + (1-p)^2 (L).
\end{align*}
The inequality becomes $ p \geq \frac{L}{N-L} $.

$ 3L \leq N $ should necessarily hold for both inequalities to hold.

Therefore, $ M \geq 2N$ and $ 3L \leq N $ is a necessary condition for attaining interference-free DoF for all $p<1$.

\subsection{$M \geq 2N$ and $N \leq L < 2N$}
{Individual DoF bound:} $ M+L > N$ and $ L \geq N $ give us
\begin{align*}
d_1,d_2 & \leq \min \left\{
p \min ( M, N+L ), N
\right\}.
\end{align*}

From $M \geq 2N$ and $N+L \geq 2N$, we have $p \min (M, N+L) \geq p(2N)$. Hence, for $\frac{1}{2} \leq p < 1$, the second term of the big minimum function is active.
\begin{align*}
d_1, d_2 \leq N.
\end{align*}

{Sum DoF bound:} $ M+L-N > N$, $L-N < N$, $ 2M + L > N $, $ M+L > N $, and $L \geq N $ give us
\begin{align*}
d_1 + d_2& \leq \min \left\{
p \min ( M-N, N+L ), p (N) + (1-p) (L-N)
\right\} + N.
\end{align*}

From $\min(a,b) \leq a$ and $\min(a,b) \leq b$, the sum DoF is bounded by
\begin{align*}
d_1 + d_2 \leq p(N) + (1-p)(L-N) + N.
\end{align*}

From $L-N < N$, we have $p(N) + (1-p)(L-N) < N$. Hence, the sum DoF is strictly tighter than the sum of the two individual DoF bounds for $\frac{1}{2} \leq p < 1$.

Therefore, $M \geq 2N$ and $N \leq L < 2N$ is not a necessary condition for attaining interference-free DoF for all $p<1$.

\subsection{$M \geq 2N$ and $L \geq 2N$}
{Individual DoF bound:} $ M+L > N$ and $ L > N $ give us
\begin{align*}
d_1,d_2 & \leq \min \left\{
p \min ( M, N+L ), N
\right\}.
\end{align*}

{Sum DoF bound:} $ M+L-N > N$, $L-N \geq N$, $ 2M + L > N $, $ M+L > N $, and $L > N $ give us
\begin{align*}
d_1 + d_2 & \leq \min \left\{
p \min ( M-N, N+L ), N
\right\} + N.
\end{align*}

Let $p_1^*$ and $p_2^*$ be the threshold probabilities that activate the second terms of the first and second big minimum functions, respectively.
\begin{align*}
p_1^* = \frac{N}{\min ( M, N+L )},\text{ } p_2^* = \frac{N}{\min ( M-N, N+L )}.
\end{align*}

When $ M - N < N + L $, $p_1^* < p_2^*$ holds. For $p_1^* \leq p < p_2^*$, the individual DoF is bounded by $N$, and the sum DoF is bounded by $p(M-N) + N$. From $p < p_2^* = \frac{N}{M-N}$, we have $p(M-N) < N$. Hence, the sum DoF bound is strictly tighter than the sum of the two individual DoF bounds.

When $ M - N \geq N + L $, the individual DoF is bounded by $\min \left\{ p(N+L), N \right\}$, and the sum DoF is bounded by $\min \left\{ p(N+L), N \right\} + N$. Hence, the sum of the two individual DoF bounds is tighter than the sum DoF bound.

Therefore, $M \geq 2N + L$ and $L \geq 2N$ is a necessary condition for attaining interference-free DoF for all $p<1$, whereas $2N \leq M < 2N + L$ and $L \geq 2N$ is not.

The necessary condition for attaining interference-free DoF in Corollary~\ref{cor:nec} is proved.

\section{Generalization of Section~\ref{subsec:412}}
\label{app:412gen}
All transmitting nodes always apply zero-forcing precoding. Each transmitter uses $2N+L$ antennas. It sends {$N$ fresh} symbols to its intended receiver, and {$L$ fresh} symbols to the relay. It uses the remaining $N$ antennas to participate in {cooperative interference nulling}. The relay uses $L$ antennas when receiving, and $2N$ antennas when sending. It uses $N$ antennas for \mbox{receiver 1} only, and the other $N$ antennas for \mbox{receiver 2} only.

Except for the number of antennas being used, there is little difference between the generalized scheme and the scheme presented in Section~\ref{subsec:412}. Since the relay is shared by both users, unavoidable collisions take place at the relay. A key idea is that the relay cooperates with active transmitters to remove the interference in the air and deliver only desired symbols to each receiver.

Until both transmitters become active, the channel can be viewed as two independent bursty relay channels.

When both transmitters become active, $L$ collisions occur at the relay. From this point, each transmitter starts to send $N$ symbols to the other receiver to participate in {cooperative interference nulling}. From feedback, the transmitters are aware of the past collisions and the order of the occurrences. In resolving the past collisions, they send $N$ symbols to the other receiver in a first-in-first-out (FIFO) manner. Also from feedback, the transmitters figure out whether or not they have succeeded in cooperative interference nulling. The relay sends its symbols adaptively for cooperative interference nulling based on current traffic states.

There is one special case to note: a series of successive occurrences of both transmitters being active. After the first occurrence, both transmitters start to send symbols to the other receiver to resolve the past collisions at the relay. This would cause interference at the receivers if both transmitters become active {again}. But the relay can prevent such interference from occuring. Let us consider an example of two successive occurrences of both transmitter being active.
\begin{itemize}
\item {Time 1}: \mbox{Transmitter 1} sends ($a_1, \cdots, a_{N+L}$), \mbox{transmitter 2} sends ($b_1, \cdots, b_{N+L}$), and the relay sends nothing. $L$ collisions ($a_1+b_1, \cdots, a_{L}+b_{L}$) occur at the shared relay. ($a_{L+1}, \cdots, a_{N+L}$) and ($b_{L+1}, \cdots, b_{N+L}$) arrive at the intended receivers without interference.
\item {Time 2}: To participate in {cooperative interference nulling}, \mbox{transmitter 1} sends ($-a_1, \cdots, -a_N$) to \mbox{receiver 2}, in addition to a new set of fresh symbols ($a'_{1}, \cdots, a'_{N+L}$) toward \mbox{receiver 1} and the relay. Also, \mbox{transmitter 2} sends ($-b_1, \cdots, -b_N$) to \mbox{receiver 1}, and ($b'_{1}, \cdots, b'_{N+L}$) toward \mbox{receiver 2} and the relay. This would cause interference: ($a'_{L+1}-b_1, \cdots, a'_{N+L}-b_N$) at \mbox{receiver 1}, and ($b'_{L+1}-a_1, \cdots, b'_{N+L}-a_N$) at \mbox{receiver 2}. But the relay knows the current traffic states, so it sends ($a_1+b_1, \cdots, a_{L}+b_{L}$) to both receivers. As a result, there is no interference at both receivers: ($a'_{L+1}+a_1, \cdots, a'_{N+L}+a_N$) at \mbox{receiver 1}, and ($b'_{L+1}+b_1, \cdots, b'_{N+L}+b_N$) at \mbox{receiver 2}. Since all desired relay-passing symbols will be eventually decoded, the new set of $N$ fresh symbols at each receiver will be also decoded.
\end{itemize}

At each receiver, there is {no interference at all times}. The relay exploits information of current traffic states, and always makes sure there is no interference at both receivers in cooperation with active transmitters. Overall, each transmitter-receiver link communicates as if there is no other link nearby.

An analysis similar to that in Section~\ref{subsec:412} is as follows.

{\bf Low-traffic Regime $p < \frac{N}{N+L}$:}
\begin{itemize}
\item Type 1 symbols: collisions. $L$ symbols are reserved at the relay with probability $p^2$, and $N$ symbols can be delivered to \mbox{receiver 1} through {cooperation} between the relay and \mbox{transmitter 2} with probability $(1-p)p$.
\begin{align*}
p^2 \times L < (1-p)p \times N.
\end{align*}

\item Type 2 symbols: collision-free. $L$ symbols are reserved at the relay with probability $p(1-p)$, and $N$ symbols can be delivered to \mbox{receiver 1} by the relay with probability $(1-p)^2$.
\begin{align*}
p(1-p) \times L < (1-p)^2 \times N.
\end{align*}

\end{itemize}

In the low-traffic regime, the above conditions hold. Each transmitter sends $N+L$ fresh symbols at the rate of $p$, and all of them will be eventually decoded at the intended receiver: the individual DoF of $p(N+L)$.

{\bf High-traffic Regime $\frac{N}{N+L} \leq p < 1$:} In this regime, both transmitters send symbols at a \emph{lower} rate. This lowering is represented as a multiplicative factor $q$.
\begin{itemize}
\item Type 1 symbols: $(pq)^2 \times L < (1-pq)(pq) \times N$.
\item Type 2 symbols: $(pq)(1-pq) \times L < (1-pq)^2 \times N$.
\end{itemize}

In the high-traffic regime, defining $q$ as $\frac{1}{p}(\frac{N}{N+L} - \epsilon)$, where $\epsilon > 0$, satisfies the above conditions. Each transmitter sends $N+L$ fresh symbols at the rate of $pq$, and all of them will be eventually decoded at the intended receiver: the individual DoF of $pq(N+L)$. As both transmitters choose $\epsilon$ arbitrarily close to zero, the individual DoF converges to $N$.
\begin{align*}
pq(N+L) = \left( \frac{N}{N+L} - \epsilon \right)(N+L) \rightarrow N.
\end{align*}

The proposed scheme achieves the following DoF region.
\begin{align*}
\mathcal{D} = \{(d_1, d_2) : d_1, d_2 \leq \min ( p(N+L), N )\}.
\end{align*}

\section{Generalization of Section~\ref{subsec:731}}
\label{app:731gen}
Both transmitters always apply zero-forcing precoding, but the relay cannot. Each transmitter uses $2N+L$ antennas. It sends $N$ fresh symbols to its intended receiver, and $L$ fresh symbols to the relay. It uses the remaining $N$ antennas to provide the other receiver with side information. The relay uses $L$ antennas when receiving and sending.

Except for the number of antennas being used, there is little difference between the generalized scheme and the scheme presented in Section~\ref{subsec:731}. Since the relay broadcasts its symbols due to its limited number of antennas, unavoidable collisions take place at the receivers. A key idea is that each transmitter provides the other receiver with side information to help the receiver resolve the interference.

Each transmitter provides the other receiver with side information at all times. It provides the duplicate of its relay-passing symbols, at most $N-L$ of them, in a first-in-first-out (FIFO) manner. From feedback, the transmitters figure out whether or not they have succeeded in providing side information. Each receiver has $N$ antennas. The receiver uses $L$ antennas to get symbols from the relay when its corresponding transmitter is inactive. And, the receiver uses $N-L$ antennas to get side information from the other transmitter if it is active.
 
The relay performs its receive-broadcast operation in a FIFO manner. When there is an inactive transmitter, the relay broadcasts the oldest $L$ symbols that have not been delivered to the corresponding receiver. This causes interference at the other receiver. When both transmitters are inactive, the relay broadcasts the $L$ sums of the oldest $L$ symbols of both users, so that both receivers get useful symbols.

An analysis similar to that in Section~\ref{subsec:731} is as follows.

{\bf Low-traffic Regime $p < \frac{1}{2}$:}
\begin{itemize}
\item User 1's relay-passing symbols: $L$ symbols are reserved with \mbox{probability $p$}, and $L$ symbols can be delivered to \mbox{receiver 1} by the relay with \mbox{probability $1-p$}.
\begin{align*}
p \times L < (1-p) \times L.
\end{align*}

\item User 2's relay-passing symbols: $L$ symbols are reserved with \mbox{probability $p$}, and eventually broadcast. \mbox{Receiver 1} can get the duplicate of $N-L$ \mbox{user 2}'s relay-passing symbols as {side information} from \mbox{transmitter 2} with probability $(1-p)p$.
\begin{align*}
p \times L < (1-p)p \times (N-L).
\end{align*}
\end{itemize}

In the low-traffic regime, the above conditions hold. Each transmitter sends $N+L$ fresh symbols at the rate of $p$, and all of them will be eventually decoded at the intended receiver in the low-traffic regime: the individual DoF of $p(N+L)$.

{\bf High-traffic Regime $\frac{1}{2} \leq p < 1$:} In this regime, both transmitters send symbols to the relay at a \emph{lower} rate. This lowering is represented as a multiplicative factor $q$.

\begin{itemize}
\item User 1's relay-passing symbols are reserved at the rate of $pq \times L$, and can be delivered at the rate of $(1-p) \times L$.
\begin{align*}
pq \times L < (1-p) \times L.
\end{align*}

\item User 2's relay-passing symbols are reserved at the rate of $pq \times L$, and eventually broadcast. \mbox{Receiver 1} can get {side information} at the rate of $(1-p)p \times (N-L)$. \begin{align*}
pq \times L < (1-p)p \times (N-L).
\end{align*}

\end{itemize}

In the high-traffic regime, defining $q$ as $\frac{1}{p}(1-p- \epsilon)$, where $\epsilon > 0$, satisfies the above conditions. Each transmitter sends $N$ fresh symbols to its intended receiver at the rate of $p$, and $L$ fresh symbols to the relay at the rate of $pq$. All of them will be eventually decoded at the intended receiver: the individual DoF of $pN + pqL$. As both transmitters choose $\epsilon$ arbitrarily close to zero, the individual DoF converges to $pN + (1-p)L$.
\begin{align*}
pN + pqL = pN + (1 - p - \epsilon)L \rightarrow pN + (1-p)L.
\end{align*}

The proposed scheme achieves the following DoF region.
\begin{align*}
\mathcal{D} = \{(d_1, d_2) : d_1, d_2 \leq pN + \min ( p, 1-p )L\}.
\end{align*}

\section{Proof of Theorem~\ref{thm:siso}}
\label{app:siso}
This appendix proves Theorem~\ref{thm:siso}. Theorem~\ref{thm:outer} proves the converse. The cut-set argument proves achievability of the individual DoF~\cite{gamalkim:nit}. We prove achievability of the sum DoF.

\textbf{Low-traffic Regime $p < \frac{1}{2}$:} The transmitters always send fresh symbols. The relay operates as follows. 

When only one transmitter is active, the relay sends nothing. The intended receiver decodes its desired symbol.
 
When both transmitters are active, the relay sends nothing. Each receiver cannot decode its desired symbol instantaneously. It gets a linear sum of its desired symbol and an undesired symbol from the other transmitter. The relay gets another linear sum that is linearly independent of the linear sum at each receiver. 

When both transmitters become inactive, the relay forwards the linear sum. Then, each receiver can decode its desired symbol that it could not decode due to interference.

In summary, both receivers get a collision of their desired symbol and an undesired symbol of the other transmitter when both transmitters are active. Both receivers decode their desired symbol in the collision when they get an extra linear sum from the relay when both transmitters become inactive.

In the low-traffic regime, it is more likely for both transmitters to be inactive ($(1-p)^2$) than active ($p^2$). In other words, the relay can forward extra linear sums to help the receivers resolve collisions more often than both receivers get collisions. The decoding of all desired symbols is guaranteed at both receivers. The sum DoF is $2p$.

\textbf{High-traffic Regime $\frac{1}{2} \leq p < 1$:} The transmitters send fresh symbols at a \emph{lower} rate to guarantee the decoding. At any time instant, each transmitter chooses to send a symbol with \mbox{probability $q$}. Each transmitter makes such decisions independently over time, and the decisions of the transmitters are independent. Hence, each transmitter is in fact active with \mbox{probability $pq$}.

In the high-traffic regime, both transmitters set $q$ to be $\frac{1 - \epsilon}{2p}$. Then, it is more likely for both transmitters to be inactive ($(1-pq)^2$) than active ($(pq)^2$). The decoding of all desired symbols is guaranteed at both receivers. The sum DoF is $2pq$. As both transmitters choose $\epsilon$ arbitrarily close to zero, the sum DoF converges to 1.

The proposed scheme achieves the following DoF region.
\begin{align*}
\mathcal{D} = \{(d_1, d_2) : d_1, d_2 \leq p, \ d_1 + d_2 \leq \min (2p, 1) \}.
\end{align*}


%
%

\ifCLASSOPTIONcaptionsoff
  \newpage
\fi



\bibliographystyle{IEEEtran}
\bibliography{bib_relayburstyic}

\begin{thebibliography}{10}
\providecommand{\url}[1]{#1}
\csname url@samestyle\endcsname
\providecommand{\newblock}{\relax}
\providecommand{\bibinfo}[2]{#2}
\providecommand{\BIBentrySTDinterwordspacing}{\spaceskip=0pt\relax}
\providecommand{\BIBentryALTinterwordstretchfactor}{4}
\providecommand{\BIBentryALTinterwordspacing}{\spaceskip=\fontdimen2\font plus
\BIBentryALTinterwordstretchfactor\fontdimen3\font minus
  \fontdimen4\font\relax}
\providecommand{\BIBforeignlanguage}[2]{{%
\expandafter\ifx\csname l@#1\endcsname\relax
\typeout{** WARNING: IEEEtran.bst: No hyphenation pattern has been}%
\typeout{** loaded for the language `#1'. Using the pattern for}%
\typeout{** the default language instead.}%
\else
\language=\csname l@#1\endcsname
\fi
#2}}
\providecommand{\BIBdecl}{\relax}
\BIBdecl

\bibitem{jafar:it09}
V.~R. Cadambe and S.~A. Jafar, ``Degrees of freedom of wireless networks with
  relays, feedback, cooperation, and full duplex operation,'' \emph{IEEE
  Transactions on Information Theory}, vol.~55, no.~5, pp. 2334--2344, May
  2009.

\bibitem{gamalkim:nit}
A.~{El Gamal} and Y.-H. Kim, \emph{Network Information Theory}.\hskip 1em plus
  0.5em minus 0.4em\relax Cambridge University Press, 2011.

\bibitem{sahin:it11}
O.~Sahin, O.~Simeone, and E.~Erkip, ``Interference channel with an out-of-band
  relay,'' \emph{IEEE Transactions on Information Theory}, vol.~57, no.~5, pp.
  2746--2764, May 2011.

\bibitem{yener:it12}
Y.~Tian and A.~Yener, ``Symmetric capacity of the {G}aussian interference
  channel with an out-of-band relay to within 1.15 bits,'' \emph{IEEE
  Transactions on Information Theory}, vol.~58, no.~8, pp. 5151--5171, Aug.
  2012.

\bibitem{sridharan:isit08}
S.~Sridharan, S.~Vishwanath, S.~A. Jafar, and S.~{Shamai (Shitz)}, ``On the
  capacity of cognitive relay assisted gaussian interference channel,''
  \emph{IEEE International Symposium on Information Theory}, Jul. 2008.

\bibitem{sahin:07}
O.~Sahin and E.~Erkip, ``Achievable rates for the {G}aussian interference relay
  channel,'' \emph{IEEE Global Telecommunications Conference}, Nov. 2007.

\bibitem{Cover:it79}
T.~M. Cover and A.~{El Gamal}, ``Capacity theorems for the relay channel,''
  \emph{IEEE Transactions on Information Theory}, vol.~25, pp. 572--584, 1979.

\bibitem{carleial}
A.~B. Carleial, ``Interference channels,'' \emph{IEEE Transactions on
  Information Theory}, vol.~24, no.~1, pp. 60--70, Jan. 1978.

\bibitem{goldsmith:it12}
I.~Mari\'c, R.~Dabora, and A.~Goldsmith, ``Relaying in the presence of
  interference: Achievable rates, interference forwarding, and outer bounds,''
  \emph{IEEE Transactions on Information Theory}, vol.~58, no.~7, pp.
  4342--4354, Jul. 2012.

\bibitem{yu:ita10}
P.~Razaghi and W.~Yu, ``Universal relaying for the interference channel,''
  \emph{IEEE Information Theory Workshop}, Feb. 2010.

\bibitem{yener:it11}
Y.~Tian and A.~Yener, ``The {G}aussian interference relay channel: Improved
  achievable rates and sum rate upperbounds using a potent relay,'' \emph{IEEE
  Transactions on Information Theory}, vol.~57, no.~5, pp. 2865--2879, May
  2011.

\bibitem{noisy:it11}
S.~H. Lim, Y.-H. Kim, A.~{El Gamal}, and S.-Y. Chung, ``Noisy network coding,''
  \emph{IEEE Transactions on Information Theory}, vol.~57, no.~5, pp.
  3132--3152, May 2011.

\bibitem{kang:isit13}
B.~Kang, S.-H. Lee, S.-Y. Chung, and C.~Suh, ``A new achievable scheme for
  interference relay channels,'' \emph{IEEE International Symposium on
  Information Theory}, Jul. 2013.

\bibitem{tobias:it14}
H.~T. Do, T.~J. Oechtering, and M.~Skoglund, ``Layered coding for the
  interference channel with a relay,'' \emph{IEEE Transactions on Information
  Theory}, vol.~60, no.~10, pp. 6154--6180, Oct. 2014.

\bibitem{wang:spawc13}
I.-H. Wang and S.~Diggavi, ``Interference channels with bursty traffic and
  delayed feedback,'' \emph{IEEE International Workshop on Signal Processing
  Advances in Wireless Communications}, Jun. 2013.

\bibitem{Katti:ACM}
S.~Katti, H.~Rahul, W.~Hu, D.~Katabi, M.~M\'edard, and J.~Crowcroft, ``{XOR}s
  in the air: Practical wireless network coding,'' \emph{IEEE/ACM Transactions
  on Networking}, vol.~16, no.~3, pp. 497--510, Jun. 2008.

\bibitem{SuhTse}
C.~Suh and D.~Tse, ``Feedback capacity of the {G}aussian interference channel
  to within 2 bits,'' \emph{IEEE Transactions on Information Theory}, vol.~57,
  no.~5, pp. 2667--2685, May 2011.

\bibitem{MaddahAli:it12}
M.~Maddah-Ali and D.~Tse, ``Completely stale transmitter channel state
  information is still very useful,'' \emph{IEEE Transactions on Information
  Theory}, no.~7, pp. 4418--4431, Jul. 2012.

\end{thebibliography}

%








\end{document}